\documentclass[twocolumn,aps,prl,superscriptaddress,showpacs,secnumroman,showkeys]{revtex4}
\usepackage{amsmath,bm}%
\usepackage{graphicx}%
\begin{document}
\title{Isobaric Yield Ratios and The Symmetry Energy In Fermi Energy Heavy Ion Reactions}  
\author{M. Huang}
%\email[E-mail at:]{huang@comp.tamu.edu}
\affiliation{Cyclotron Institute, Texas A$\&$M University, College Station, Texas 77843}
\affiliation{Institute of Modern Physics, Chinese Academy of Sciences, Lanzhou, 730000,China.}
\author{Z. Chen}
\affiliation{Cyclotron Institute, Texas A$\&$M University, College Station, Texas 77843}
\affiliation{Institute of Modern Physics, Chinese Academy of Sciences, Lanzhou, 730000,China.}
\author{S. Kowalski}
\affiliation{Institute of Physics, Silesia University, Katowice, Poland.}
\author{Y.G. Ma}
\affiliation{Shanghai Institute of Applied Physics, Chinese Academy of Sciences, Shanghai, 201800,China}
\author{R. Wada}
\email[E-mail at:]{wada@comp.tamu.edu}
\affiliation{Cyclotron Institute, Texas A$\&$M University, College Station, Texas 77843}
\author{T. Keutgen}
\affiliation{FNRS and IPN, Universit\'e Catholique de Louvain, B-1348 Louvain-Neuve, Belgium}
\author{K. Hagel}
\affiliation{Cyclotron Institute, Texas A$\&$M University, College Station, Texas 77843}
\author{J. Wang}
\affiliation{Institute of Modern Physics, Chinese Academy of Sciences, Lanzhou, 730000,China.}
\author{L. Qin}
\affiliation{Cyclotron Institute, Texas A$\&$M University, College Station, Texas 77843}
\author{J. B. Natowitz}
\affiliation{Cyclotron Institute, Texas A$\&$M University, College Station, Texas 77843}
\author{T. Materna}
\affiliation{Cyclotron Institute, Texas A$\&$M University, College Station, Texas 77843}
\author{P.K. Sahu}
\affiliation{Cyclotron Institute, Texas A$\&$M University, College Station, Texas 77843}
\author{M.Barbui}
\affiliation{Cyclotron Institute, Texas A$\&$M University, College Station, Texas 77843}
\author{C.Bottosso}
\affiliation{Cyclotron Institute, Texas A$\&$M University, College Station, Texas 77843}
\author{M.R.D.Rodrigues}
\affiliation{Cyclotron Institute, Texas A$\&$M University, College Station, Texas 77843}
\author{A. Bonasera}
\affiliation{Cyclotron Institute, Texas A$\&$M University, College Station, Texas 77843}
\affiliation{Laboratori Nazionali del Sud, INFN,via Santa Sofia, 62, 95123 Catania, Italy}

\date{\today}

\begin{abstract}
The relative isobaric yields of fragments produced in a series of heavy ion induced multi-fragmentation reactions have been analyzed in the framework of a Modified Fisher Model, primarily to determine the ratio of the symmetry energy coefficient to the temperature, $a_{a}/T$, as a function of fragment mass A. The extracted values increase from 5 to $\sim$ 16 as  A increases from 9 to 37. These values have been compared to the results of calculations using the Antisymmetrized Molecular Dynamics (AMD) model together with the statistical decay code Gemini. The calculated ratios  are in good agreement with those extracted from the experiment. In contrast, the ratios determined from fitting the primary fragment distributions from the AMD model calculation are $\sim$ 4 and show little variation with A.  This observation indicates that the value of the symmetry energy coefficient derived from final fragment observables may be significantly  different than the actual value at the tim
 e of fragment formation.The experimentally observed pairing effect is also studied within the same simulations. The Coulomb coefficient is also discussed.  
\end{abstract}
 


\maketitle
 
\section*{I. INTRODUCTION}

In the early 1980's, a study of the isotopic yield distributions of intermediate mass fragments produced in high energy proton induced multi-fragmentation reactions at Fermi Lab showed that the distributions can be well described by a Modified Fisher Model~\cite{Minich82,Hirsch84} in which the isotope production is governed by the available free energy. Therefore isotope yields provide a good probe to study the nature of the disassembling nuclear system. Multi-fragmentation of the system is also generally observed in violent collisions in heavy ion reactions in the Fermi energy domain and there is evidence that both subnormal and supernormal densities may be explored in such collisions~\cite{Kowalski07,BALi03}. Work in this area has concentrated on exploring the nuclear equation of state and the liquid gas phase transition in nuclear matter. In a recent paper we addressed the possibility of probing the quantum nature of the liquid-gas phase transition using isotope distributi
 ons~\cite{Bonasera08}.  

Over the last several years many fragment emission studies have been motivated by efforts to use fragment yield distributions, either singly or by comparison to those of  similar reactions, to explore the symmetry energy in the emitting source at different densities and temperatures~\cite{Kowalski07,HSXu00,Tsang01,Botvina02,Ono03}. In each of these cases  measuring the isotope distributions over a wide range of mass number A and atomic number Z should provide a more reliable basis for extraction of the desired information. Even then there are important issues which must be resolved in order to establish the relation between the experimental isotope distributions and the symmetry energy of the emitting system. One is the source temperature T. In the Modified Fisher Model as well as other approaches based upon the free energy, all terms which can be determined from experiments appear in the form of $a_{i}/T$ times some function of A, Z or the neutron number N, where i indicates
  the coefficient of the different terms contributing to  the free energy. Since the beginning of the experimental study of heavy ion collisions in the multifragmentation regime, significant efforts have been made to evaluate the source temperature, but no absolute consensus among different methods has yet been achieved~\cite{Kelic06}. A second issue is the role and effect of secondary decay processes. In experiments the majority of the detected fragments are in their ground states.  Most of the primary fragments produced in Fermi energy heavy ion reactions are expected to be in an excited state when they are formed. Indeed, in previous work, excitation energies of the primary fragments have been evaluated by studying the associated light charged particle multiplicities~\cite{Marie98,Hudan03}. Such data demonstrate that secondary decay is important and raise the question of the degree of confidence which can be accorded to experimental derivations of  the symmetry energy coef
 ficient which do not properly correct for this important effect. A clear goal for experimentalists would be the reconstruction of the primary isotope distributions from the experimental distributions. This might be approached by a reconstruction of the primary isotope distributions employing the associated neutron and charged particle multiplicities. However, since multiple fragments are produced in a reaction and light particles can be produced even before the formation of the fragments, the identification of the parent for a detected light particle observed in coincidence with detected fragments is not straight forward. A reconstruction of the primary isotope distribution was part of goal of the experiment described here, but that analysis is still underway~\cite{Wada05}.

A third issue, not addressed in this paper but of critical importance in this field, is the problem of obtaining reliable estimates of the density at the time of fragment formation. In the absence of such information most density estimates should be viewed as unconfirmed. While this may be somewhat mitigated by comparing experimental observables with results of dynamic models employing  particular assumed forms for the density dependence of the symmetry energy, such approaches are integral and may be influenced by other assumptions and parameter choices  inherent in the model applied~\cite{Moretto08}.

In this paper we explore the extent to which information on the symmetry energy, in the form of $a_{a}/T$,  can be extracted from high quality data for isotope resolved fragment yield distributions and compared to the model 
predictions~\cite{Ono03,Raduta07}. The role of the secondary decay is explored by comparisons with results of theoretical calculations.  In a future paper we will discuss the extraction of such information using isoscaling techniques~\cite{Chen}.
  
\section*{II. EXPERIMENT}

\begin{figure}
\includegraphics[scale=0.40]{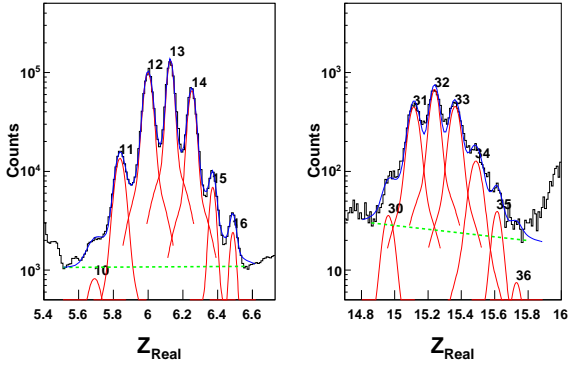}
\caption{\footnotesize Typical linearized isotope spectra for Z=6 and 15 are shown for $^{64}$Ni+$^{124}$Sn. The number at the top of each peak is the assigned mass number. The histograms depict experimental data. A linear back ground is assumed from valley to valley for a given Z. Each isotope is fit with two Gaussians. The individual fit indicates the yield of the isotope above the back ground. The sum of Gaussians and the background are also shown in each spectrum. 
}
\label{fig:fig_1}
\end{figure}

The experiment was performed at the K-500 superconducting cyclotron facility at Texas A$\&$M University. $^{64,70}$Zn and $^{64}$Ni beams were used to irradiate $^{58,64}$Ni, $^{112,124}$Sn, $^{197}$Au and $^{232}$Th targets at 40 A MeV. Intermediate mass fragments (IMFs) were detected by a detector telescope placed at 20$^\circ$. The telescope consisted of four Si detectors. Each Si detector was 5cm x 5cm. The nominal thicknesses were 129, 300, 1000, 1000 $\mu$m. All Si detectors were segmented into four sections and each quadrant had a 5$^\circ$ opening angle in polar and azimuthal angles. Therefore the energies of the fragments were measured at two polar angles of the quadrant detector, namely $\theta$ = 17.5$^\circ$ $\pm$ 2.5$^\circ$ and $\theta$ = 22.5$^\circ$ $\pm$ 2.5$^\circ$. Typically 6-8 isotopes for atomic numbers, Z, up to Z=18 were clearly identified with the energy threshold of 4-10 A MeV, using the $\Delta$E-E technique for any two consecutive detectors. The $\
 Delta$E-E spectrum was linearized  empirically. Mass identification of the isotopes were made using a range-energy table~\cite{Hubert90}. In the analysis code, isotopes are identified by a parameter $Z_{Real}$. For the isotope with A=2Z, $Z_{Real}$ = Z is assigned and other isotopes are identified by interpolation  between them. Typical $Z_{Real}$ spectra are shown in Fig.\ref{fig:fig_1}. The energy spectrum of each isotope was extracted by gating the isotope in a 2D plot of $Z_{Real}$ vs energy. The yields of light charged particles (LCPs) in coincidence with IMFs were also measured using 16 single crystal CsI(Tl) detectors of 3cm thickness  set around the target. The light output from each detector was read by a photo multiplier tube. The pulse shape discrimination method was used to identify p, d, t, h and $\alpha$ particles. The energy calibration for these particles were performed using Si detectors (50 -300 $\mu$m) in front of the CsI detectors in separate runs.

The yield of each isotope was evaluated, using a moving source fit. For LCPs, three sources (projectile-like(PLF), nucleon-nucleon-like(NN) and target-like (TLF)) were used. The NN-like sources have source velocities of about a half of the beam velocity. The parameters are searched globally for all 16 angles. For IMFs, since the energy spectra were measured only at the two angles of the quadrant detector, the spectra were parameterized using a single NN-source. Using a source with a smeared source velocity around  half the beam velocity, the fitting parameters were first determined from the spectrum summed over all isotopes for a given Z, assuming A=2Z. Then all extracted parameters except for the normalizing yield parameter were used for the individual isotopes. This procedure was based on the assumption that, when the spectrum is plotted in energy per nucleon, the shape of the energy spectrum is same for all isotopes for a given Z. Indeed the observed energy spectra of isot
 opes are well reproduced by this 
method. For IMFs, a further correction was made for the background.
For the IMFs, a further correction was  made for the background. As seen in Fig.\ref{fig:fig_1}, the isotopes away from the stability line, such as $^{10}$C and $^{36}$P, have a very small yields and the background contribution is significant. In order to evaluate the background contribution to the extracted yield from the source fit, a two Gaussian fit to each isotope combined with a linear background was used. The fits are shown in Fig.\ref{fig:fig_1}. Each peak consists of two Gaussians. The second Gaussian (about 10\% of the height of the first one) is added to reproduce the shape of the valley between two isotopes. This component is attributed to the reactions of the isotope in the Si detector. The centroid of the Gaussians was set to the value calculated from the range-energy table within a small margin. The final yield of an isotope with Z $>$ 2 was determined  by correcting the yield evaluated from the moving source fit by the ratio between the two Gaussian yields and
  the linear background.
Rather large errors ( $\sim \pm10\%$) are assigned for the multiplicity of the NN source for IMFs, originating 
from the source fit besides the background estimation. The errors from the source fit are evaluated from the different assumptions of the parameter set for the source velocity and temperature.

\section*{III. MODIFIED FISHER MODEL}

In order to study the symmetry energy contribution to the isotope production, the Modified Fisher Model of ref.1 is used. In the Modified Fisher model, the fragment yield of A nucleons with I=N-Z, Y(A,I) is given by
\begin{eqnarray}
Y(A,I) &=& CA^{-\tau}exp\{[(W(A,I)+\mu_{n}N+ \mu_{p}Z)/T] \nonumber\\
&&+Nln(N/A)+Zln(Z/A)\}.     
\label{eq:eq_1}
\end{eqnarray}
C is a constant. The A$^{-\tau}$ term originates from the entropy of the fragment and the last two terms are from the entropy contributions for the mixing of two substances in the Fisher Droplet Model~\cite{Fisher67}. $\mu_{n}$ is the neutron chemical potential and  $\mu_{p}$ is the proton chemical potential. W(A,I) is the free energy of the cluster at temperature T. As such it includes both energy and entropy terms. In the model W(A,I) is given by the following generalized Weisz$\ddot{a}$cker-Beth semi-classical mass formula~\cite{Weizsacker35,Bethe36} at a given temperature T and density $\rho$,
\begin{eqnarray}
W(A,I) &=& - a_{a}(\rho,T)I^{2}/A-a_{c}(\rho,T)Z(Z-1)/A^{1/3} \nonumber\\
&&+a_{v}(\rho,T)A- a_{s}(\rho,T)A^{2/3}- \delta(N,Z). 
\label{eq:eq_2}
\end{eqnarray}
The indexes, v, s, c and a represent volume, surface, Coulomb, and symmetry energy, respectively. Following the semi-empirical mass formulation,  the pairing energy, $\delta$(N,Z), is  given by~\cite{Green53,Preston62}
\begin{eqnarray}
\delta(N,Z)=\left\{\begin{array}{ll} 
 a_{p}(\rho,T)/A^{1/2 }&(\textrm{odd-odd})\\
 0 &(\textrm{even-odd})\\       
-~a_{p}(\rho,T)/A^{1/2 }&(\textrm{even-even} ).
\end{array}\right.
\label{eq:eq_3}
\end{eqnarray}
We define the isotope yield ratio, R(I+2,I,A), between isobars differing by 2 units in I as
\begin{widetext}
\begin{equation*}
R(I+2,I,A) = Y(A,I+2)/Y(A,I)
\label{eq:doubleraitio1}
\end{equation*}
\begin{equation}
           =exp\{[(W(I+2,A)-W(I,A)+ (\mu_{n}- \mu_{p})]/T + S_{mix} (I+2,A) -  S_{mix} (I,A)\},
\label{eq:eq_4}
\end{equation}
\end{widetext}
where $S_{mix}(I,A) = Nln(N/A) + Z ln(Z/A)$. Hereafter, in order to simplify the description, the density and temperature dependence of the coefficients in Eq.(\ref{eq:eq_2}) is omitted as a$_{i}$= a$_{i}$($\rho$,T) (i=v,s,c,a,p). Inserting Eq.(\ref{eq:eq_2}) into Eq.(\ref{eq:eq_4}), one can get 
\begin{widetext}
\begin{equation}
R(I+2,I,A) =  exp\{[\mu_{n}- \mu_{p}+ 2a_{c}(Z-1)/A^{1/3}- 4a_{a}(I+1)/A
                   -\delta(N+1,Z-1) - \delta(N,Z)]/T+ \Delta(I+2,I,A)\},
\label{eq:eq_5}
\end{equation}
\end{widetext}
where $\Delta(I+2,I,A)=S_{mix}(I+2,A) - S_{mix}(I,A)$. One should note that $\Delta(1,-1,A) = 0$ and for other I values $\Delta(I+2,I,A)\leq 0.5$, which is rather small comparing other parameters in Eq.(\ref{eq:eq_5}).
\begin{figure}[ht]
\includegraphics[scale=0.35,clip]{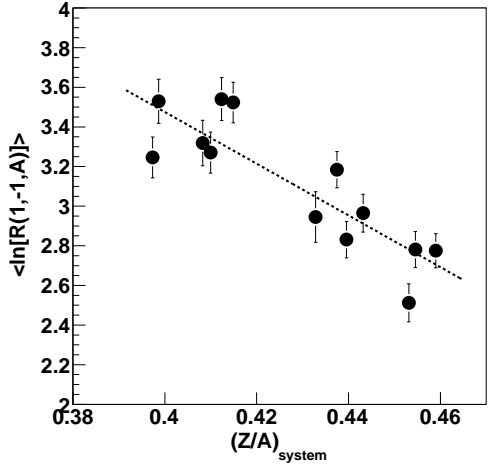}
\caption{\footnotesize Experimental average values of ln[R(I+2,I,A)] for the case of I=-1 are plotted as a function of Z/A of the reaction systems. Z/A = (Z$_{p}$+Z$_{t}$)/(A$_{p}$+A$_{t}$) where p and t represent the projectile and the target, respectively. The dotted line is a linear fit.  
}
\label{fig:fig_2}
\end{figure}
                                       
\begin{figure}
\includegraphics[scale=0.35]{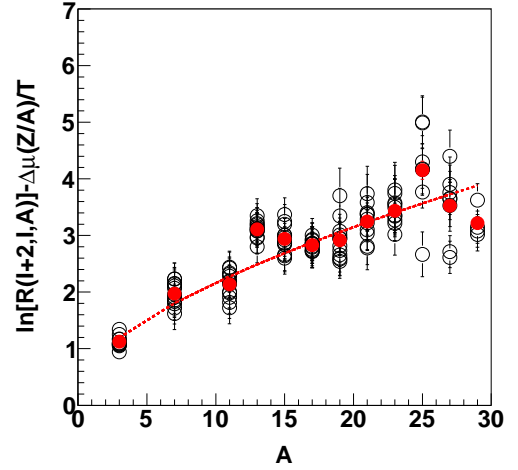}
 \caption{\footnotesize Experimental values of ln[R(1,-1,A)] with the offset correction for different reactions is plotted as a function of A for I=-1. Open circles show results  from the individual experiments and solid circles depict the average values for a given A over all reactions. The dotted line shows the result of fitting the average values with Eq.(\ref{eq:eq_6}).
}
\label{fig:fig_3}
\end{figure}

Initially we focus on the isobars with I=-1 and 1.  For these isobars the contributions from the symmetry term and the mixing entropy term in Eq.(\ref{eq:eq_5}) drop out and, since these isobars are even-odd nuclei, the pairing term  also drops out. Taking the logarithm of the resultant equation, one can get
\begin{equation}  
ln[R(1,-1,A)] = [(\mu_{n}- \mu_{p}) + 2a_{c}(Z-1)/A^{1/3}]/T.
\label{eq:eq_6}
\end{equation}
For different reaction systems, N/Z and thus $(\mu_{n}- \mu_{p})/T$ can be different. In order to evaluate the system dependence of $(\mu_{n}- \mu_{p})/T$ we determined the average value of ln[R(1,-1,A)] over all available fragments for each system. In Figure \ref{fig:fig_2} these average values are plotted as a function of the entrance channel Z/A for the reaction systems studied. As seen in the figure, the average values show a linear dependence on the entrance channel Z/A of the system. Since the Coulomb energy in the right hand side of Eq.(\ref{eq:eq_6}) is that of the fragment itself and therefore expected to be very similar for the different reaction systems and the temperature T is also assumed similar because the same incident energy is used, we attribute the linear dependence seen in the figure to the difference of  $(\mu_{n}- \mu_{p})/T$ in the different systems. Expressing $(\mu_{n}- \mu_{p})/T$ as
\begin{equation}
([\mu_{n}- \mu_{p})/T]_i = [(\mu_{n}- \mu_{p})/T]_0 + \Delta\mu(Z/A)/T\nonumber, 
\label{eq:ChePo1}
\end{equation}
and
\begin{equation}
\Delta\mu(Z/A)/T = c1\cdot (Z/A) + c2,
\label{eq:eq_7}
\end{equation}
here $[\mu_{n}- \mu_{p})/T]_i$ denotes the value for a given reaction system i, and $[(\mu_{n}- \mu_{p})/T]_0$ is for the reference reaction system.  A linear fit to the expression (\ref{eq:eq_7}) gives c1=-13.0 and c2=8.7 for Fig.(\ref{fig:fig_2}). We took the $^{64}$Zn+$^{112}$Sn reaction as the reference, i.e., the extracted values in the figure have been adjusted to the reference reaction using $\Delta\mu(Z/A)/T$, in which $\Delta\mu(Z/A)/T$=0 for the $^{64}$Zn+$^{112}$Sn reaction. In Fig.\ref{fig:fig_3} the experimental values of ln[R(1,-1,A)] corrected by $\Delta\mu$(Z/A)/T are plotted for all reactions as a function of A. Fitting  these corrected average values using $(\mu_{n}- \mu_{p})/T$ and $a_{c}/T$ as fitting parameters, Eq.(\ref{eq:eq_6}) leads to $(\mu_{n}- \mu_{p})/T$=0.71  and $a_{c}/T$=0.35.
                                  
\begin{figure}
\includegraphics[scale=0.35]{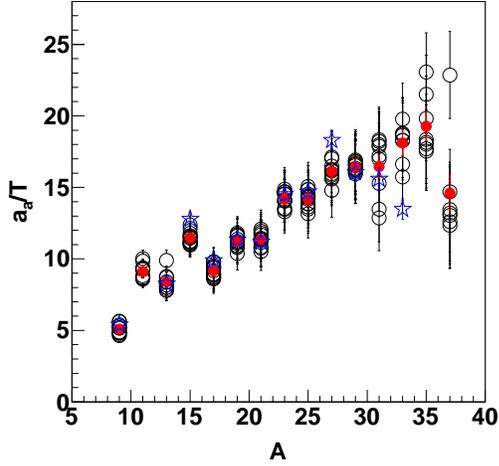}
 \caption{\footnotesize Experimental values of $a_{a}/T$ as a function of A.  Open circles are obtained from Eq.(\ref{eq:eq_8}) and solid circles are the average values for a given A.  Stars are the average values obtained from Eq.(8'). 
}
\label{fig:fig_4}
\end{figure}

We next compare  isobars with I=1 and 3, noting that these isobars are also even-odd nuclei for which the pairing term is 0. For this combination, the symmetry energy coefficient  term in Eq.(\ref{eq:eq_5}) is given as a function of A by
\begin{eqnarray}
a_{a}/T= -A/8\{ln[R(3,1,A)] - [(\mu_{n}- \mu_{p})/T \nonumber\\
      + 2a_{c}(Z-1)/A^{1/3}]/T - \Delta\mu(Z/A)/T- \Delta(3,1,A) \}.	
\label{eq:eq_8}
\end{eqnarray}
In Fig.\ref{fig:fig_4} values of $a_{a}/T$ calculated from Eq.(\ref{eq:eq_8}) using the values $(\mu_{n}- \mu_{p})/T$=0.71 and $a_{c}/T$=0.35 determined above, are plotted as a function of A. All available values from the different reactions are plotted in the figure. The extracted values are very similar in magnitude and trend for the different reactions. In general the values increase from 5 to $\sim$ 16 as A increases from 9 to 27 and may show a plateauing above that. 

\begin{figure}
\includegraphics[scale=0.35]{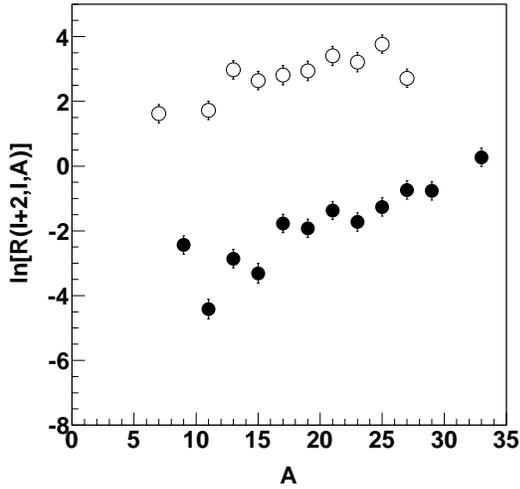}
\caption{\footnotesize Experimental values of ln[R(I+2.I,A)] for I=-1(open circles) and I=1(solid circles) for the $^{64}$Zn+$^{112}$Sn reaction.
}			
\label{fig:fig_5}
\end{figure}

The symmetry term can also be extracted without evaluating the values of  $(\mu_{n}- \mu_{p})/T$ and $a_{c}/T$ explicitly. In Fig.\ref{fig:fig_5}, the experimental values of ln[R(3,1,A)] and ln[R(1,-1,A)] from the $^{64}$Zn+$^{112}$Sn reaction are plotted. The symmetry term, $a_{a}/T$, for a give A can be extracted approximately by the difference of these values as
\begin{eqnarray*}
a_{a}/T \sim  -A/8\{ln[R(3,1,A)] - ln[R(1,-1,A)]\\
- \Delta(3,1,A) \}.\hspace{0cm}      ~(8')
\label{eq:eq_8s}
\end{eqnarray*}
The approximation made in Eq.(8') is that the Coulomb term in ln[R(3,1,A)] is same as that in ln[R(1,-1,A)]. In the actual calculation, the Coulomb term in ln[R(3,1,A)] for A=13, for example, is calculated from the yield ratio of $^{13}$B/$^{13}$C, whereas that in ln[R(1,-1,A)] is calculated from $^{11}$B/$^{11}$C, assuming that the ratio of the Coulomb energy for $^{13}$B/$^{13}$C is same as that of $^{11}$B/$^{11}$C. A similar approximation has been made for all extracted values. The resultant symmetry values are plotted in Fig.\ref{fig:fig_4} using star symbols. The main difference between the values from Eq.(\ref{eq:eq_8}) (circles) and those from Eq.(8') (stars) originates from the deviation of the data from the fitted line in Fig.\ref{fig:fig_3}. 
               
The pairing terms in Eq.(\ref{eq:eq_5}) can be determined using the extracted values for $(\mu_{n}- \mu_{p})/T$, $a_{c}/T$ and $a_{a}/T$ for the combination of isobars with I=0 and 2 and with I=2 and 4. For I=0 and 2 isobars, the pairing term can be written as
\begin{eqnarray}
	a_{p}/T = (sign)(1/2)A^{1/2}\{ln[R(2,0,A)]- [ (\mu_{n}- \mu_{p})\nonumber\\
                +2a_{c}(Z-1)/A^{1/3}  - 4a_{a}/A ]/T  - \Delta(2,0,A) \},
\label{eq:eq_9}
\end{eqnarray}
and for I=2 and 4 it is given by
\begin{eqnarray}
	a_{p}/T = (sign)(1/2)A^{1/2}\{ln[R(4,2,A)]- [ (\mu_{n}- \mu_{p})\nonumber\\ 
                 + 2a_{c}(Z-1)/A^{1/3}  - 12a_{a}/A]/T - \Delta(4,2,A) \}.\hspace{0.1cm}
\label{eq:eq_10}
\end{eqnarray}
Here sign=1 for (N,Z)=(odd,odd) and -1 for (even,even) nucleus, in which A=N+Z.
\begin{figure}
\includegraphics[scale=0.35]{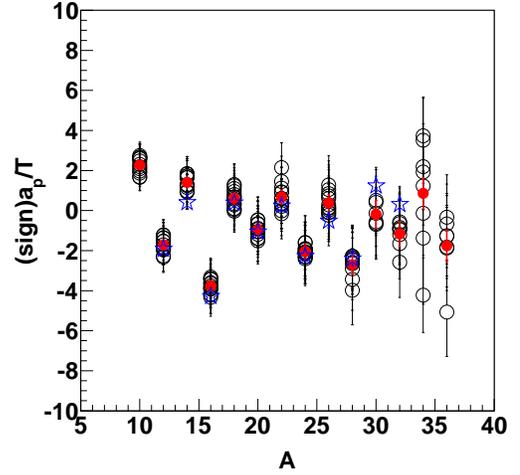}
\includegraphics[scale=0.35]{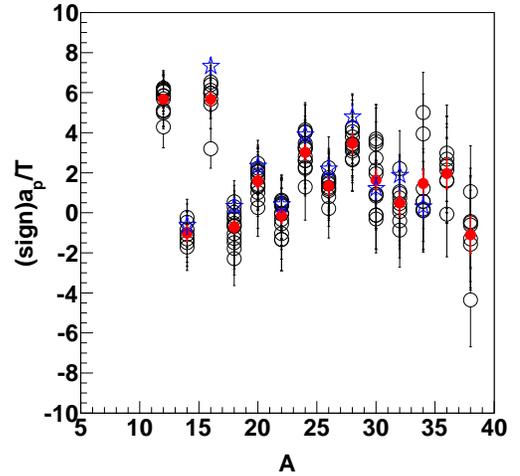}
\caption{\footnotesize Extracted values  of (sign)$a_{p}/T$ are plotted as a function of A. Open circles are obtained, using Eq.(\ref{eq:eq_9}) on the upper and Eq.(\ref{eq:eq_10}) on the lower, respectively, for individual reactions. Solid circles are averaged values for a given A over all reactions. Stars are obtained from Eqs.(\ref{eq:eq_11}) and (\ref{eq:eq_12}).
}
\label{fig:fig_6}
\end{figure}
The $a_{p}/T$ values obtained from Eqs.(\ref{eq:eq_9}) and (\ref{eq:eq_10}), using the extracted values of $(\mu_{n}- \mu_{p})/T$, $a_{c}/T$ and $a_{a}/T$ above, are plotted in Fig.\ref{fig:fig_6}. On the top for I=0 and 2 isobars, the pairing contribution is clearly observed for isobars with A $<$ 30, though the even-odd oscillation pattern is slightly distorted. On the other hand on the bottom, using Eq.(\ref{eq:eq_10}) with I=2 and 4 isobars, only clear pairing effect is observed for isotopes only with A $<$ 20.
 
	The pairing term, $a_{p}/T$, can also be extracted from experimental yield ratios of isobars without the explicit evaluation of $(\mu_{n}- \mu_{p})/T$, $a_{c}/T$ and $a_{a}/T$, similar to Eq.(8').  Inserting Eq.(\ref{eq:eq_6}) and (8') into Eq.(\ref{eq:eq_9}) with I=0 and 2 isobars, one can get 
\begin{widetext}
\begin{eqnarray}
a_{p}/T  \sim 	(sign)(1/2)A^{1/2}\{ ln[R(2,0,A)]-ln[R(1,-1,A)] - 1/2(ln[R(3,1,A)] - ln[R(1,-1,A)] \nonumber\\
- \Delta(3,1,A))- \Delta(2,0,A)\}\nonumber\\
= (sign)(1/2)A^{1/2}\{ln[R(2,0,A)]-1/2(ln[R(1,-1,A)] + ln[R(3,1,A)]\nonumber\\
-\Delta(3,1,A))- \Delta(2,0,A)\}. 
\label{eq:eq_11}
\end{eqnarray}
\end{widetext}
From Eq.(\ref{eq:eq_10}) with I=2 and 4 isobars, 
\begin{widetext}
\begin{eqnarray}
a_{p}/T  \sim (sign)(1/2)A^{1/2}\{ln[R(4,2,A)]-ln[R(1,-1,A)] - 3/2(ln[R(3,1,A)] - ln[R(1,-1,A)]\nonumber\\ 
-\Delta(3,1,A))- \Delta(4,2,A)\}\nonumber\\
= (sign)(1/2)A^{1/2}\{ ln[R(4,2,A)] +1/2( ln[R(1,-1,A)] - 3ln[R(3,1,A)]\nonumber\\
+3\Delta(3,1,A))- \Delta(4,2,A)\}.
\label{eq:eq_12}
\end{eqnarray}
\end{widetext}           
The resultant values are plotted in Fig.\ref{fig:fig_6} using star symbols. These are the averaged values over all reactions. The results are consistent to those from Eqs.(\ref{eq:eq_9}) and (\ref{eq:eq_10}).

\section*{IV.	COMPARISONS WITH MODEL CALCULATIONS}

	In the multi-fragmentation regime of heavy ion reactions, fragments may be formed in excited states~\cite{Marie98,Hudan03}. Such fragments will de-excite by statistical decay processes. The experimentally detected fragments are normally in the ground state. In order to study the effect of the secondary decay process on the experimentally extracted symmetry energy coefficient, we have used an Antisymmetrized Molecular Dynamics (AMD) Code ~\cite{Ono03,Ono96,Ono99} to model  the reaction dynamics and coupled it with the statistical decay code Gemini~\cite{Charity88} to model the secondary decay processes. The AMD code has previously been used to study the fragment production in Fermi energy heavy ion reactions and the global features of the experimental results have been well reproduced~\cite{Ono02,Wada98,Ono04,Wada00,Wada04,Hudan06}. We believe that the dynamics  can be crucial in accounting for fragment production in early stages of the reaction and the use of a dynamic model
 , such as AMD is essential.                                   
\begin{figure}
\includegraphics[scale=0.35]{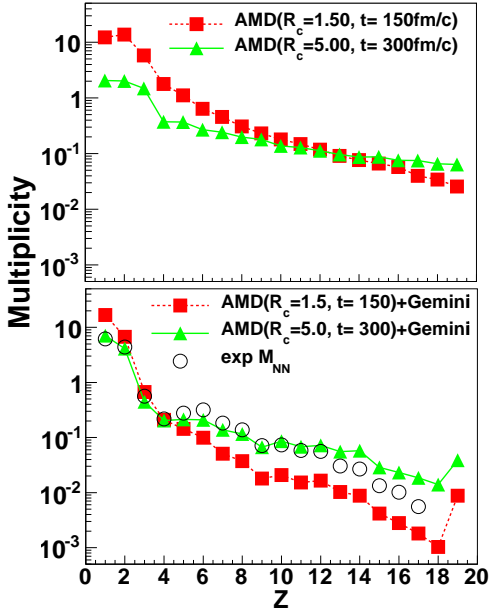}
 \caption{\footnotesize (Upper) Calculated multiplicity distributions of the primary fragments evaluated  using $R_{c}$=1.5 and 5. See details in the text. (Lower) Multiplicity distributions of fragments for the experiments and the calculations. The experimental values are shown by open circles. The AMD+Gemini calculated values filtered by the experimental acceptance are shown for $R_{c}$=1.5 (squares) and $R_{c}$=5 (triangles).  All errors evaluated are smaller than the size of the symbols.
}
\label{fig:fig_7}
\end{figure}

Since the AMD calculation requires a lot of CPU time, only two of the experimental reaction systems have been studied in detail. The systems examined were $^{64}$Zn+$^{112}$Sn and $^{64}$Ni+$^{124}$Sn, both at 40 A MeV. All results shown in this paper have been calculated using a newly installed computer cluster in the Cyclotron Institute~\cite{Wada09}. The calculations have been performed, using the Gogny interaction with an asymptotic stiff symmetry energy term~\cite{Ono03}, although very similar results are obtained for the standard Gogny interaction. In order to obtain yields of the final products, the yields of primary fragments are first evaluated at a given time in the offline analysis. The fragments are formed using a coalescence radius, $R_{c}$, in phase space. Because the hot and dense composite system formed at an early stage of the reaction expands very quickly, the primary fragment distributions are rather sensitive to the choice of coalescence radius and the tim
 e of its application When a smaller $R_{c}$ is used, one can form fragments at an earlier stage. In  previous calculations, $R_{c}$= 5  and t=300fm/c were used and the experimental results were well reproduced~\cite{Wada98,Wada00,Wada04}. $R_{c}$=5 corresponds to a radius of 5 fm in configuration space. In order to study the effect of the choice of these parameters, two different coalescence radii, $R_{c}$ = 1.5 and 5, are used here. In the case of $R_{c}$=5, the fragment formation is   evaluated at t = 300fm/c. For $R_{c}$ = 1.5, the evaluation is at t = 150fm/c. The excitation energy of a fragment is calculated by subtracting the binding energy from the total energy. For each isotope the binding energy is calculated within the AMD code using a stochastic cooling method~\cite{Ono92}. In the upper part of Fig.\ref{fig:fig_7}, the primary fragment distributions are shown as a function of the fragment Z for the two different cases. As one can see, for $R_{c}$ = 1.5, the multip
 licity of light IMFs with Z $<$ 10 is significantly enhanced, compared to that for $R_{c}$=5.0, whereas the heavier fragment yields are suppressed as expected. The deexcitation of these primary  fragments was followed using the Gemini code~\cite{Charity88}. The Gemini code has been used extensively with the AMD simulations in the past and good agreement  with the experimental data has been seen~\cite{Wada98,Wada00,Wada04}. In order to sample all possible decay channels of the excited fragments, one AMD event is used 100 times in Gemini, with different random number seeds. In the lower part of Fig.\ref{fig:fig_7} the multiplicity distributions of the secondary fragments are compared with the experimental values. Since the experimental values are taken from the NN-source component of the energy spectra, the calculated energy spectra are subjected to the same filter. For simplicity the calculated range of the impact parameter is set to 0 $<$ b $<$ 8 fm to suppress the contribut
 ion from the projectile like fragments from the peripheral collisions. The impact parameter range was determined from the correlation between the collision centrality and  impact parameter studied in ref.30. The angle range, determined from the extracted moving source parameters, was set to 5$^\circ$ $<$ $\theta$ $<$ 45$^\circ$ to suppress the heavy projectile-like contribution and target-like contributions.  In addition the experimental energy threshold is applied for detection of each isotope. The experimentally observed multiplicity distribution for most particles, including Z=1 and 2, is well reproduced by the $R_{c}$= 5 calculation. When $R_{c}$=1.5 is applied, the multiplicity is significantly overestimated for Z=1 and 2 and underestimated for Z $>$ 4. In the case of $R_{c}$=5, the calculated multiplicities start to deviate from those of the experiments at Z $>$ 13, indicating that the projectile like contribution becomes significant for these fragments.               
                     
\begin{figure}
\includegraphics[scale=0.35]{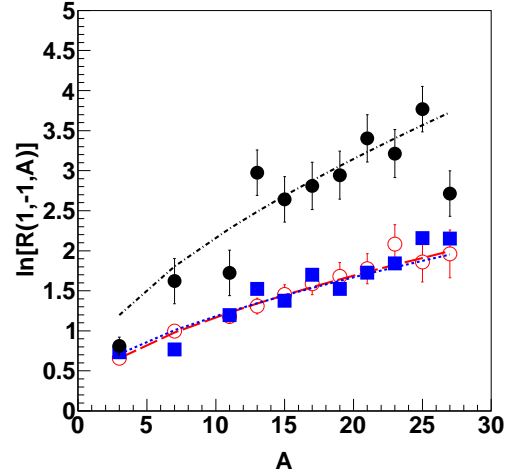}  
\caption{\footnotesize ln[R(I+2,I,A)] for I=-1 and 1 primary (open circles) and secondary (filled squares) fragments of AMD-Gemini events for the  $^{64}$Zn+$^{112}$Sn reaction. $\Delta\mu$(Z/A)/T=0 for this reaction. Solid circles are the experimental values for this reaction. Dotted, dashed and Dashed-dotted lines show the fits, using Eq.(\ref{eq:eq_6}) for the experiments, the primary and the secondary fragments, respectively. 
}
\label{fig:fig_8}
\end{figure}

Using the same prescription used for the experimental data, $a_{c}/T$, $a_{a}/T$ and $a_{p}/T$ are evaluated from the filtered yields of isobars. In Fig.\ref{fig:fig_8}, ln[R(I+2,I,A)] for the isobars with I=-1 and 1 is plotted as a function of A for both primary and  secondary fragments, together with the experimental results. No notable difference is observed between these two set of the calculated fragments. The calculated results are also well fitted by Eq.(\ref{eq:eq_6}) and the values for $(\mu_{n}- \mu_{p})/T$ and $a_{c}/T$ are extracted from the distributions of the primary and secondary isobars. The extracted values are $(\mu_{n}- \mu_{p})/T$=0.40 and $a_{c}/T$=0.18 for the primary isobars and $(\mu_{n}- \mu_{p})/T$=0.47 and $a_{c}/T$=0.17 for those of the secondary( Similar results are obtained from the unfiltered yields). The values following secondary decay should be compared to the experimental values of $(\mu_{n}- \mu_{p})/T$=0.71 and $a_{c}/T$=0.35. The calcula
 ted values of $(\mu_{n}- \mu_{p})/T$ both from the primary and secondary fragments are somewhat lower than that of the experiments, as seen in the figure. The Coulomb coefficient for the calculations is about a half of the experimental value, which suggests that the fragments in the calculations are more expanded and/or deformed than those of the experiments. 
\begin{figure}
\includegraphics[scale=0.35]{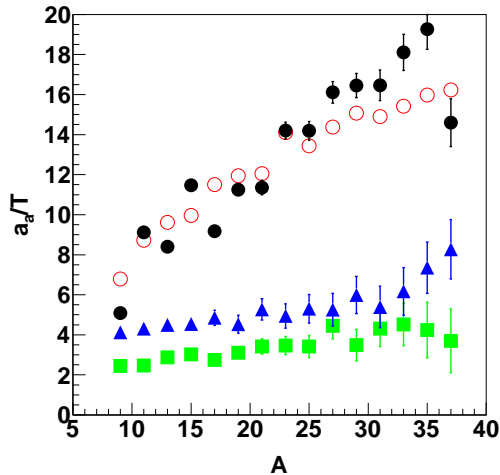}  
\caption{\footnotesize Extracted values of symmetry energy coefficient from the experiments (solid circles) and calculations from the secondary fragments for $R_{c}$=5 (circles). Squares and triangles show those obtained for primary fragments for $R_{c}$=1.5 and $R_{c}$=5, respectively. All errors evaluated are smaller than the size of the symbols. 
}
\label{fig:fig_9}
\end{figure}

In Fig.\ref{fig:fig_9}, $a_{a}/T$ values evaluated using Eq.(\ref{eq:eq_8}) with the values of $(\mu_{n}- \mu_{p})/T$ and $a_{c}/T$ extracted from the model calculations are plotted as a function of A. The average values of the experimental results from Fig.\ref{fig:fig_4} (solid circles) may be compared with those extracted following the de-excitation step in the Gemini Code(open circles). Although the experimental values show larger fluctuations, the calculated values and the data are in good agreement in magnitude and trend. In the figure the values extracted from the primary fragment yields are also plotted for both $R_{c}$ = 1.5 (squares) and 5 (triangles). The values for $R_{c}$=5 show a slowly increasing distribution as a function of A with $a_{a}/T$ $\sim$ 4 to 5 over most of the range. Those for $R_{c}$=1.5 show a similar trend, but the values are smaller. The  rather flat distribution of $a_{a}/T$ values of the primary fragments for A $<$ 30 is consistent with a sce
 nario of fragment emission from a common  source with a given density and temperature. The heavier fragments with A $>$ 30 may result from different mechanisms, e.g., projectile fragmentation in the more peripheral collisions.
\begin{figure}
\includegraphics[scale=0.35]{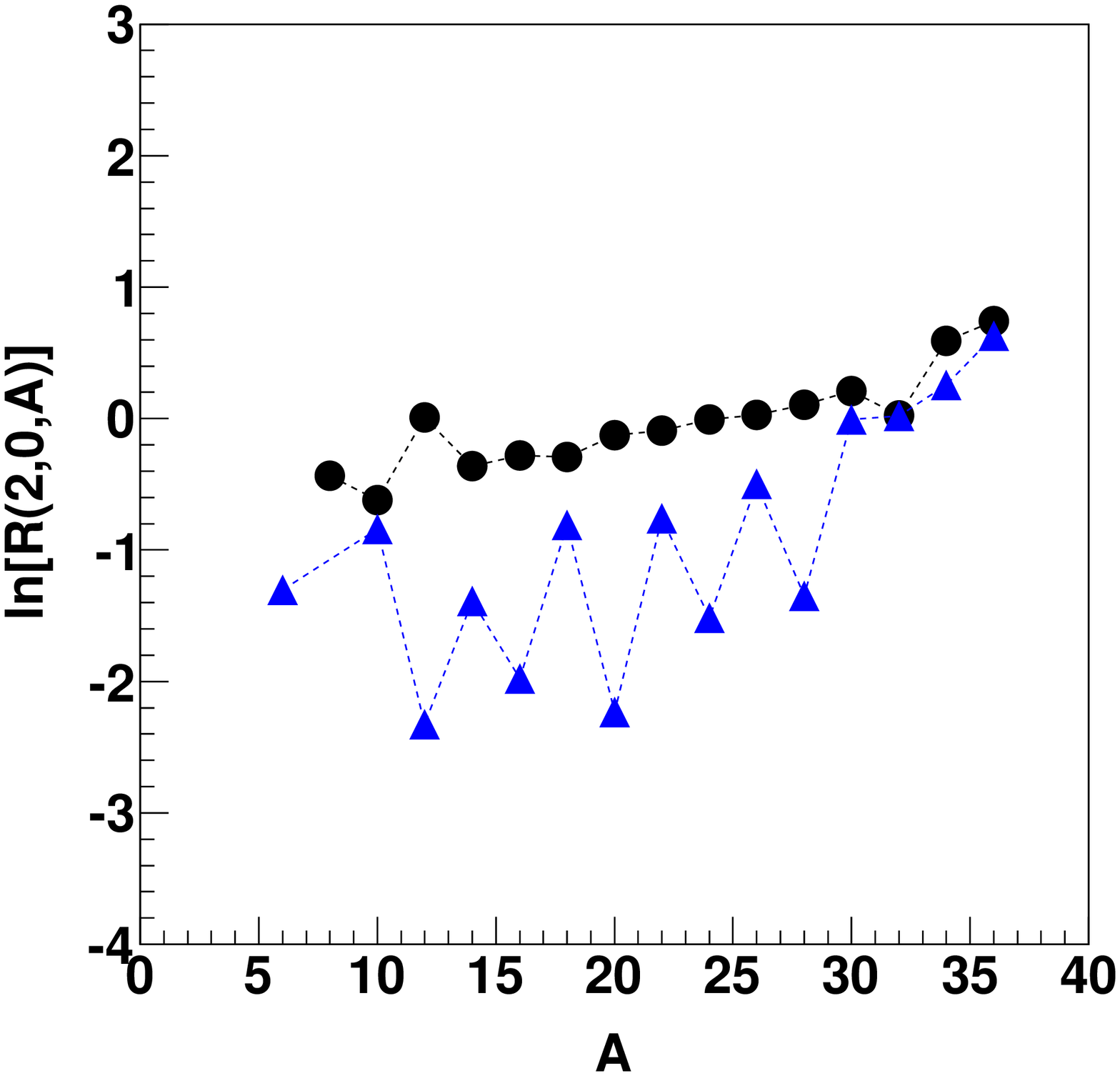}
\includegraphics[scale=0.35]{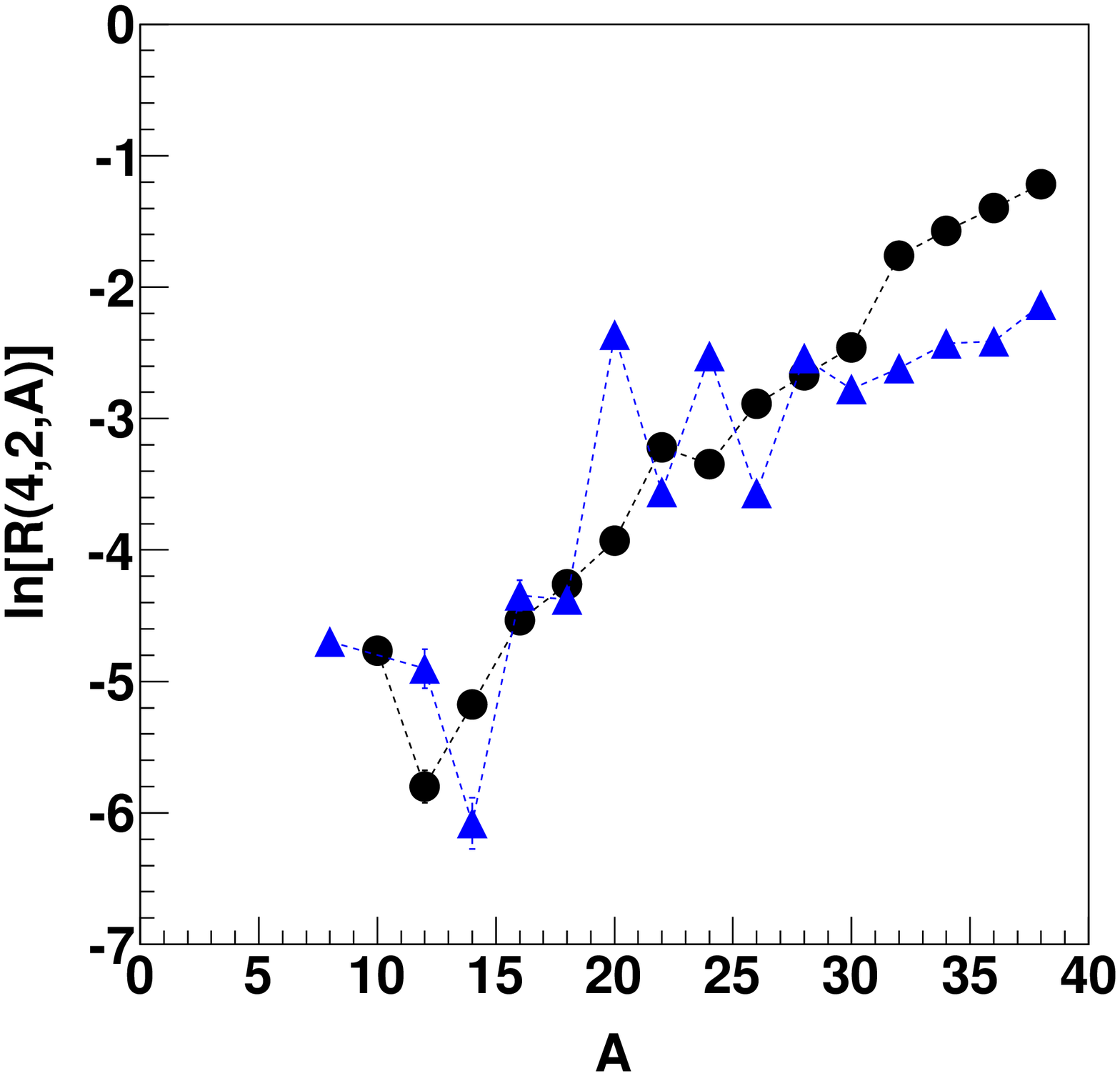}    
\caption{\footnotesize Calculated ln[R(I+2,I,A)] values are plotted for I=0 in the upper panel and for I=2 in the lower panel.  Triangles show results for fragments in the ground states, solid circles are for those in excited states just before the last particle decay. Dotted lines are to guide the eye. 
}
\label{fig:fig_10}
\end{figure}

The pairing effect is experimentally observed in Fig.\ref{fig:fig_6}. In the study of the complex fragments production in the $^{238}$U + Ti reaction at 1 A GeV, Ricciardi et al. suggested that the observed even-odd oscillation in the fragment yield originates from the last chance particle decay of the excited fragments during their cooling down evaporation process~\cite{Ricciardi04,Ricciardi05}. In order to verify this hypothesis in our experiment, we also did a similar study, using AMD-Gemini calculations. Two sets of simulated events were prepared. In one of the set, all fragments were in the ground state. In the other set, the last chance particle decay was blocked and all fragments were in an excited state just before the last chance particle decay. In the actual calculations, the second set was generated from the first set by adding the charge and mass of the last chance emitted particle to that of the partner IMF. 
In Fig.\ref{fig:fig_10}, the results calculated for the different fragment sets are shown for ln[R(2,0,A)] and ln[R(4,2,A)] values in Eqs.(\ref{eq:eq_9}) and (\ref{eq:eq_10}), in which the pairing effect reveals, if it is sufficiently large. Calculated values of ln[R(2,0,A)] are shown in the upper and  those of ln[R(4,2,A)] are in the lower figure. 
As one can see in the upper figure for the case of isobars with I=0 and 2, the values calculated for the fragments in the ground state show a clear even-odd oscillation pattern and the patterns are completely washed out for fragments observed just before the last particle decay. This is consistent with the experimental observation shown in the upper part of Fig.\ref{fig:fig_6}. Even the disappearance of oscillations above A=30 is well reproduced. The same feature is also observed for isobars with I=2 and 4, but the oscillation pattern is more distorted. In this case the results for the fragments in the ground state show the oscillatory pattern up to A = 28, but the pattern is not as clear as those in the experiments, especially for A $<$ 20, which are shown in the lower part of Fig.\ref{fig:fig_6}. The oscillatory pattern is totally smeared out for the fragments at an excited state just before the last chance particle emission, as in the I=0 case. These results are consistent
  with those reported by Ricciardi et al. in their experiments and simulations, in which an enhanced oscillation pattern is observed for I=0 isotopes and the pattern becomes less clear for other isotope distributions~\cite{Ricciardi04,Ricciardi05}. Therefore our analysis strongly supports their hypothesis that the experimentally observed pairing effects in Fig.\ref{fig:fig_6} indeed originate from the last chance particle decay during the statistical cooling down process of the excited fragments.

\section*{V.	SUMMARY}

For a large series of heavy ion reactions, coefficients of Coulomb energy, symmetry energy and pairing energy in the form of $a_{i}/T$ as a function of A have been studied from analyses of the yield ratios of isobars obtained in experiments and from model calculations.  The AMD and Gemini codes were used for the calculations. For the symmetry energy term, the extracted values from the experiments are in good agreement with those calculated for the final fragments in the ground state. They increase from 5 to $\sim$ 16 as the masses of the fragments increase from 9 to 37. These values are generally much larger than those extracted from the primary fragments observed in the AMD calculations. Over the same mass interval the primary fragment values range from 4 to 5. This is consistent with a picture in which the primary fragments originate from a common emitting source. A smaller coalescence radius and earlier sampling time for the fragment formation in the AMD calculation result
 s in slightly smaller values of 2 to 3 for the primary fragments, but does not have a strong effect on the extracted values for the final fragments, though the isotope yield distribution is not well reproduced compared to the case of $R_{c}$ = 5. Although the technique employed is quite different, our model results are quite similar to those observed by Ono where the ratio $\zeta$(Z) =C$_{sym}$/T was extracted from the calculated isotope distributions for the reactions  $^{40}$Ca+$^{40}$Ca, $^{48}$Ca+$^{48}$Ca, $^{60}$Ca+$^{60}$Ca and $^{46}$Fe+$^{46}$Fe at 35 A MeV~\cite{Ono05}. In that paper the ratios are evaluated from the quadratic shapes of the isotope distributions for a given Z after proper  normalization of the isotope yields from the three different reactions. The comparisons between the experimentally extracted results and those of the calculations indicate that the experimental determination of symmetry energy coefficients, $a_{a}/T$, are significantly affected b
 y the secondary decay processes of the primary fragments. This modification is a common feature of dynamic transport model approaches~\cite{Ono05,Liu04}. Thus extraction of the density dependence of the symmetry energy from fragment observables must be done with caution and with appropriate attention to the role of the secondary decay. The importance of these effects will vary according to the observables employed for extraction of the desired information.

The pairing effect is clearly observed in the experiments. Comparisons to the calculations strongly support the hypothesis, which is proposed by Ricciardi et al.~\cite{Ricciardi04,Ricciardi05}, that the observed effect originates from the last chance particle decay during the statistical cooling down process of the excited fragments.

The Coulomb coefficient in the form of $a_{c}/T$ is also evaluated both in the experiment and calculations. The experimentally extracted value is $a_{c}/T$ = 0.35, whereas the calculated values are $a_{c}/T$ = 0.17 for the final fragments and $a_{c}/T$ = 0.18 for the primary fragments. These difference suggest that the calculated fragments are more deformed and/or expanded than those observed in the experiments. 

\section*{ACKNOWLEGDEMENT} 

We thank the staff of the Texas A$\&$M Cyclotron facility for their support 
during the experiment. We also thank L. Sobotka for letting us to use his spherical scattering chamber. We further thank Dr. A. Ono and Dr. R. J. Charity for providing us their simulation codes.  This work was
supported by the U.S. Department of Energy under Grant No. 
DE-FG03-93ER40773 and the Robert Welch Foundation under Grant A0330. One of us(Z. Chen) also thanks the \textquotedblleft100 
Persons Project" of the Chinese Academy of Sciences for the support.

\end{document}